\documentclass[%
 reprint,
superscriptaddress,
 amsmath,amssymb,
 aps,
]{revtex4-2}

\usepackage{amssymb,amstext,amsmath}
\usepackage{graphicx}
\usepackage{dcolumn}
\usepackage{bm} 
\usepackage{bbm} 
\usepackage{multirow}
\usepackage{bigstrut}
\usepackage{makecell,rotating}
\usepackage{mathrsfs}
\usepackage{booktabs}
\usepackage{threeparttable}
\usepackage{subfigure}
\usepackage{chngpage}
\usepackage{float}
\usepackage{color}
\usepackage{mathtools}
\usepackage{url}
\definecolor{darkred}{rgb}{0.75,0,0}
\definecolor{darkgreen}{rgb}{0,0.5,0}
\definecolor{darkblue}{rgb}{0,0,0.75}
\definecolor{darkorange}{rgb}{0.8,0.3,0}
\definecolor{dark}{rgb}{0,0,0}

\begin{document}

\preprint{APS/123-QED}

\title{Personalised strategy of allocating social goods in structured populations
}
\author{Yao Meng}
\affiliation{%
Center for Systems and Control, College of Engineering, Peking University, Beijing 100871, China}
\author{Sean P. Cornelius}
\affiliation{%
Department of Physics, Toronto Metropolitan University, Toronto ON M5B 2K3, Canada}
\author{Yang-Yu Liu}%
\affiliation{%
Channing Division of Network Medicine, Department of Medicine, Brigham and Women's Hospital and Harvard Medical School, Boston, MA 02115, USA}
\author{Aming Li}
\thanks{Corresponding author: amingli@pku.edu.cn}
\affiliation{%
Center for Systems and Control, College of Engineering, Peking University, Beijing 100871, China}
\affiliation{
Center for Multi-Agent Research, Institute for Artificial Intelligence, Peking University, Beijing 100871, China}

\date{\today}

\begin{abstract}
	Cooperation underlies many aspects of the evolution of human and animal societies, where cooperators produce social goods to benefit others.
	Explaining the emergence of cooperation among selfish individuals has become a major research interest in evolutionary dynamics. Previous studies typically use complex networks to capture the interactions between individuals, and assume that cooperators distribute benefits equally to their neighbors.
	In practice, the distribution of social goods is often non-uniform, and individuals may selectively provide benefits to those they interact with based on their personal preferences.
	Here, we develop an efficient algorithm to optimize the placement of donation structure in any given network to minimize the threshold for the emergence of cooperation.
	We find when cooperators allocate the benefits preferentially compared to the traditional settings of donating to all neighbors, cooperation tends to be maximally promoted.
	Furthermore, the optimal donation structure is strongly disassortative---the low-degree nodes tend to donate to high-degree ones preferentially and vice versa.
	Based on this finding, we offer a local heuristic strategy based on degree thresholds for personalizing the allocation of social goods and choosing each cooperator's recipient,  which we use to prove its effectiveness in empirical datasets.
	Our findings advance the understanding of mechanisms for promoting cooperation with strategic allocations of social goods.
\end{abstract}

\maketitle

\section{Introduction}
Cooperative behavior, which involves incurring personal costs to benefit others, reduces the survival chances of cooperators but offers an evolutionary advantage to their selfish opponents.
Explaining how cooperators compete with self-interested individuals and become dominant has long been a central topic in modern science \cite{Nowak92Nature,hofbauer1998evolutionary,Sigmund2010,levin2020collective,hilbe2018evolution}.
There is a large body of research proposing the different mechanisms that facilitate cooperation in social dilemmas, yet there has been a lack of research on how to maximize cooperation in these scenarios.
In the classic social dilemma, a cooperator produces benefits for others, while defectors pay no costs and produce no goods.
It is natural to ask: how should the benefits produced by cooperators be distributed in societies, and does there exist an optimal distribution to promote collective cooperation?

Previous studies on the evolutionary game theory have shown that population structures---which are captured by complex networks---have a profound effect on the evolution of cooperation.
Here, nodes represent individuals and edges capture their interpersonal interactions \cite{hauert2004spatial,lieberman2005evolutionary,santos2005scale,Traulsen2005,ohtsuki2006simple,Ohtsuki2007symmetry,taylor2007evolution,Santos2008Social,tarnita2009strategy,szolnoki2009topology,perc2010coevolutionary,Sigmund2010,Szolnoki2012,perc2013evolutionary, maciejewski2014evolutionary,allen2017evolutionary,Su2019,fotouhi2019evolution,levin2020collective,McAvoy2020,li2020evolution}.
In a typical setting, each cooperator provides an equal amount of benefit to each of its neighbors with a total cost proportional to the number of neighbors \cite{ohtsuki2006simple}.
This requires a large pool of wealth for the high-degree or hub nodes with a large number of connections to pay for all their neighbors in heterogeneous structures.
In contrast, in a fixed cost and uniform benefit model, the cooperator pays a fixed cost and divides the benefit equally among its neighbors \cite{McAvoy2020}. In this setting, cooperative behaviors are generally easier to evolve compared to the traditional scenario with proportional costs and proportional benefits \cite{ohtsuki2006simple}.
All these studies have assumed that cooperators distribute benefits uniformly to all neighbors without preference.

In practice, however, individual preferences are commonly embedded in societies, which can lead to a non-uniform distribution of benefits and asymmetric social interactions \cite{hauser2019social,Su2022,McAvoy2015Asym}.
People may give favors to those they trust while withholding favors from those they dislike.
On the other hand, preferential social interactions are also abundant in populations of non-human species.
For example, worker bees provide royal jelly only to specific larvae chosen to become a queen bee \cite{knecht1990patterns}.
Therefore, such heterogeneous or preferential social resource allocation which is ubiquitous in both human and animal societies, should be incorporated into frameworks for studying the evolution of cooperation.

In this study, we explore preferential allocations of social goods in structured populations, where individuals can arbitrarily distribute social goods according to their preferences.
We propose an efficient algorithm to minimize the critical benefit-to-cost ratio for the emergence of cooperation by optimizing the preferential recipient for each individual in any structured population.
Furthermore, we offer a simple rule to further prove the effectiveness of our optimization results: nodes with degrees below a degree threshold should choose their high-degree neighbors as recipients and vice versa.
\begin{figure*}[!ht]
	\centering
	\includegraphics[width= \textwidth]{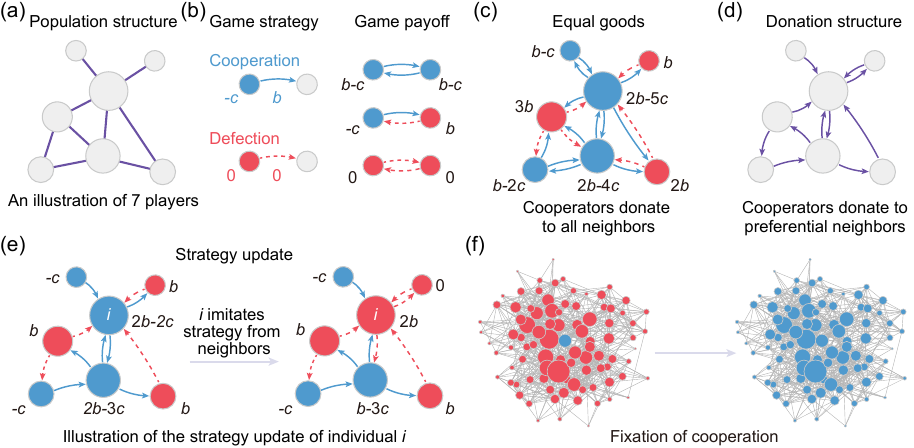}
	\caption{Illustration of evolutionary games with preferential allocations of social goods.
		(a) The network structure captures interactions between individuals, where nodes represent individuals and any two individuals play games if an edge exists between them.
		(b) Individuals choose either cooperation or defection as their game strategy.
		A cooperator provides a benefit $b$ to its opponent (blue solid arrow) by paying a cost $c$, while a defector provide no benefit (red dashed arrow) and pays no cost.
		(c) Traditionally, a cooperator provides each of its neighbors a benefit $b$, with a total cost proportional to the number of its neighbors.
		(d) In contrast, here we allow each individual have their own preferential recipients, where directed edges indicate cooperators provide a benefit $b$ to its selected neighbors.
		(e) After each game round, a random individual $i$ is chosen to update its strategy by imitating the behavior of a neighbor.
		(f) Starting from a single cooperator, the evolutionary process on a network ends whenever a state of either all cooperators or all defectors is reached.
	} \label{Fig1}
\end{figure*}

\section{Model}

We consider evolutionary game dynamics on a structured population of $N$ individuals, where interactions between individuals are represented by an undirected unweighted network (Fig.~\ref{Fig1}(a)).
Individuals choose to either cooperate or defect.
Specifically, a cooperator pays a cost ($c$) to provide a benefit ($b$) to the opponent, while defectors pay nothing and thus provide no benefit (Fig.~\ref{Fig1}(b)).
In traditional settings, a cooperator should benefit all its neighbors with a total cost proportional to the number of its neighbors (Fig.~\ref{Fig1}(c)).

The \textit{population structure} captured by the undirected network defines who can interact and imitate the strategy from whom.
In contrast, the allocation of social goods is represented by a directed network, where a directed edge $I_{ij}=1$ represents that individual $i$---when it chooses cooperation---provides a benefit to one of its neighbor $j$ preferentially over the evolutionary process, otherwise it provides no benefit to $j$ ($I_{ij}=0$), and incurs no cost.
In other words, we utilize a directed \textit{donation structure} to capture who donates to whom, and we allow each node to have its own preferential recipients for benefits (Fig.~\ref{Fig1}(d)).
In each game round, cooperators provide each such neighbor with a benefit $b$, paying a cost $c$ for each.
This naturally results in the non-uniform distribution of social goods in structured populations.
For simplicity we assume that the preferential recipients for each individual do not change throughout the game.
After tabulating the payoffs for each node, we choose one individual $i$ uniformly at random to imitate the strategy of one of its neighbors $j$ on the \textit{population structure} with probability proportional to the neighbor's fitness (Fig.~\ref{Fig1}(e)).
The fitness of $j$ is defined by $F_j = 1+ \delta f_j$, representing the ability of its strategies to be imitated, where $f_j$ captures the accumulated payoff for $j$.
Here $\delta>0$ captures the intensity of selection, which we assume to be relatively weak ($\delta \ll 1$).

To quantify the ability of cooperation to proliferate, we seed our simulations with a single cooperator placed uniformly at random in a population of defectors.
We then simulate the evolutionary game described above until a state with either all cooperators or all defectors is reached.
The fixation probability of cooperation ($\rho_\text{C}$) is the probability of reaching a state of full cooperation in this process (Fig.~\ref{Fig1}(f)).
Note that under neutral drift ($\delta=0$), the fixation probability of cooperation (defection) is $1/N$, namely the interactions have no effect on the evolutionary process, as all nodes have the same fitness ($F_i=1$).
Here we study the condition under which cooperation is favored, namely $\rho_\text{C}>1/N$ when the benefit-to-cost ratio $b/c$ exceeds the critical threshold $C^*$.

\begin{figure*}[!ht]
	\centering
	\includegraphics[width=\textwidth]{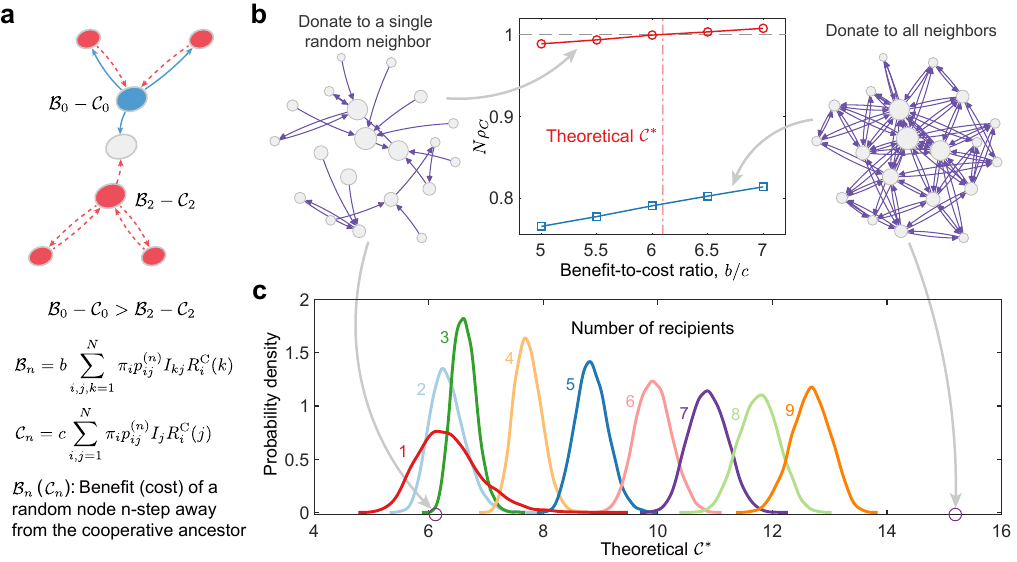}
	\caption{Effect of preferential donation structures on the emergence of cooperation.
	(a) We illustrate the key mechanism behind the evolutionary success of cooperation---namely, that cooperators are favored if they have a higher payoff than a random two-step-away individual.
	Over the course of evolution, the cooperator accumulates benefits $\mathcal{B}_0$ by paying a total cost $\mathcal{C}_0$, and the random individual two steps away obtains the payoff $\mathcal{B}_2-\mathcal{C}_2$ on average.
	(b) We show the fixation probability of cooperation ($\rho_\text{C}$)  as a function of the benefit-to-cost ratio ($b/c$) with only a single random recipient for each individual (red circle), compared with the traditional scenario where cooperators donate benefits to all their neighbors (blue square).
	The critical benefit-to-cost ratio $\mathcal{C}^*$ for the preferential donations occurs when the corresponding curve intersects the horizontal line ($\rho_\text{C}=1/N$) representing the neutral-drift case, and the theoretical $\mathcal{C}^*$ are marked with vertical (red dashed) line in (b) and purple circles in (c).
	Theoretical $\mathcal{C}^*$ with $10^4$ realizations of the randomization for preferential structures of social goods with different numbers of recipients are shown in (c), where the single recipient tend to have the smallest values.
	Numerical values of $\rho_\text{C}$ are obtained from the fraction of simulations in which the population reaches full cooperation out of $10^7$ independent realizations on networks of $N=20$ nodes generated by Barab\'asi-Albert model, with an average degree $\langle k \rangle = 6$.
	} \label{Fig2-random}
\end{figure*}

\section{Results}
\subsection{Preferential social interactions promote the emergence of cooperation}
We first derived a closed-form expression for the critical benefit-to-cost ratio $\mathcal{C}^*$ on combination of population/donation structure.
The brief idea is that, the strategy transimission occurs on the undirected population structure.
But it is the donation structure that determines how payoffs are distributed, and hence how appealing different node's strategies are.
We consider the expected payoffs of a cooperator and a random two-step neighbor over the course of the evolutionary process. Intuitively, if a cooperator obtains a higher payoff than the individual two steps away, 
then when their common neighbors update strategies, the cooperator will become more competitive in spreading its copies (Fig.~\ref{Fig2-random}(a)).
Based on this intuition, we define $\mathcal{B}_n$ ($\mathcal{C}_n$) as the benefit received (or the cost incurred) by a random node $n$-steps away from the initial cooperator over the course of evolution.

Consider an initial cooperator is placed with probability $\pi_i$, which indicates the reproductive value \cite{Fisher1930,taylor1990allele,maciejewski2014reproductive} of individual $i$.
Then a random node $j$ is reached by an $n$-step random walk away from $i$ with probability $p_{ij}^{(n)}$, who receives a benefit from its neighbor $k$ on the condition that $k$ chooses $j$ as its recipient ($I_{kj}=1$), and $k$ share the same strategy with the initial cooperator through the strategy dispersal.
We define $R_i^{\text{C}}(k)=R-\eta_{ik}$ as the total time of choosing cooperation for $k$ given node $i$ is the initial cooperator,
where $R$ is the total round of the evolutionary process, and $\eta_{ik}$ captures the time for $i$ and $k$ tracing back to a common strategy on the undirected population structure, namely the coalescence time \cite{cox1989coalescing} for two random walkers starting from $i$ and $k$, respectively.
Therefore, the expected benefit $\mathcal{B}_n = b\sum_{i,j,k} \pi_i p_{ij}^{(n)} I_{kj} R_i^{\text{C}}(k)$ is directly obtained.
Analogously, the corresponding accumulated cost paid by the individual $n$-step away from the initial cooperator is $\mathcal{C}_n = c\sum_{i,j} \pi_i p_{ij}^{(n)} I_{j} R_i^{\text{C}}(j)$, where $I_j=\sum_k I_{jk}$ indicates the number of recipients for $j$, which incurs the proportional costs paid when $j$ is a cooperator.
Therefore, cooperation is favored over defection if and only if
\begin{equation}
	\mathcal{B}_0-\mathcal{C}_0 > \mathcal{B}_2-\mathcal{C}_2,
	\label{intuition}
\end{equation}
where the $\mathcal{B}_0-\mathcal{C}_0$ represents the total payoff of the cooperator, and analogously, $\mathcal{B}_2-\mathcal{C}_2$ captures the total payoff of a random individual two steps away from the cooperator over the course of evolution.
If the cooperator has a higher payoff, it will be more capable of spreading the strategy to its neighbors compared to the random node two steps away.
As a consequence of Eq.~(\ref{intuition}), we obtain the critical benefit-to-cost ratio $\mathcal{C}^*$ above which cooperation is favored:
\begin{equation}
	\mathcal{C}^* = \frac{\sum_{i,j}\pi_i p_{ij}^{(2)}I_j \eta_{ij}}{\sum_{i,j,k}\pi_i p_{ij}^{(2)}I_{kj} \eta_{ik} - \sum_{i,j} \pi_i I_{ji} \eta_{ji} },
	\label{exact}
\end{equation}
The formal proof of this result is shown in Supplementary Information.

Based on the above theory, together with numerical simulations, we investigate how preferential donation structures affect the fate of cooperators on a networks constructed by the Barab\'asi-Albert model \cite{Barabasi1999a}.
We first allow each individual to choose a single random neighbor as its recipient (Fig.~\ref{Fig2-random}(b)).
We show that cooperation is greatly promoted compared to the traditional case where cooperators donate to all neighbors (Fig.~\ref{Fig2-random}(b)).
In this scenario, the high-degree nodes end up bearing too many costs, making it difficult for them to spread cooperation.
We then calculate the theoretical $\mathcal{C}^*$ with different numbers of random recipients and present the probability distribution of $\mathcal{C}^*$ in Fig.~\ref{Fig2-random}(c).
Surprisingly, a \textit{single} preferential recipient per cooperator tends to result in the lowest $\mathcal{C}^*$.
In contrast, and the traditional case of donating to all neighbors demands the largest $\mathcal{C}^*$.
This motivates us to first study the extreme case where each individual has a single recipient, saving the more general case for later.

\begin{figure*}[!ht]
	\centering
	\includegraphics[width=\textwidth]{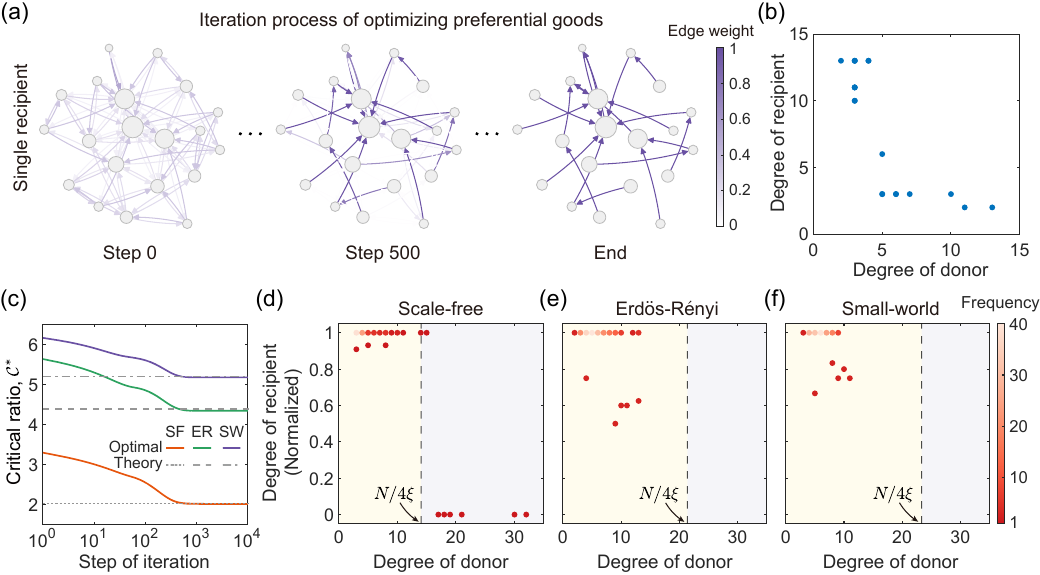}
	\caption{Optimizing recipient for each individual on heterogeneous networks.
	(a) We illustrate the network structure of preferential social goods with $N=20$ nodes at iteration step $0$, step $500$ and the end (step $7410$) of the optimization process, where each individual is allowed to have only one recipient among all neighbors.
	The color of the edges in donation structure during iteration represents the weight of the benefit.
	(b) We find that the degree of each node's recipient becomes negatively correlated with donor's degree under the optimal configuration of preferential social goods. 
	(c) For large networks, we present the convergence of the objective function $\mathcal{C}^*$ for scale-free (SF, orange line), Erd\H{o}s-R\'enyi \cite{Erdos1959} (ER, green line), and Watts-Strogatz \cite{watts1998collective} networks (WS, purple line) over $10^4$ iterations of our optimization protocol.
	The values of $\mathcal{C}^*$ corresponding to our heuristic rule are marked as gray dashed lines.
	(d-f) We show the degree of the recipient as a function of the degree of donor in the optimal configuration on SF, ER and WS networks.
	The degree of recipient is normalized to $[0, 1]$, indicating the minimal and maximal degree of neighbors, and the color of dots represents the frequency of the producer-recipient pairs.
	Under our heuristic rule to promote cooperation, nodes with degree smaller than $N/4 \xi$ (light yellow region) should choose the highest-degree neighbor as its recipient, otherwise it should choose the lowest-degree neighbor as the recipient (light purple region).
	}\label{Fig3-opt}
\end{figure*}

\subsection{Optimizing preferential allocations of social goods}
Given any population structure, can we find the optimal donation structure, namely the best preferential recipient for each individual to promote cooperation?
Here we employ a protocol based on the RMSprop \cite{tieleman2012lecture} algorithm to determine the single preferential recipient for each individual that minimizes $\mathcal{C}^*$, via iterative gradient descent.
Figure~\ref{Fig3-opt}a presents the optimal process of $\mathcal{C}^*$ for a network using the Barab\'asi-Albert model, starting from dividing a benefit equally to all neighbors.
The optimal donation structure presents significant degree dissassortivity between recipients and donors.
Specifically, the high-degree nodes tend to donate to low-degree neighbors and vice versa (Fig.~\ref{Fig3-opt}(b)).
For large networks capturing three commonly studied population structures---Watts-Strogatz \cite{watts1998collective}, Erd\H{o}s-R\'enyi \cite{erdHos1960evolution}, and scale-free \cite{Barabasi1999a} networks---we show that scale-free networks can attain a much smaller $C^*$ after optimizing recipients due to its high heterogeneity (Fig.~\ref{Fig3-opt}(c)).
Moreover, we find that low-degree nodes tend to choose their high-degree neighbors as their recipients, while high-degree nodes tend to choose the smallest ones on different network structures (Fig.~\ref{Fig3-opt}(d)-(f)).
The accuracy of our proposed algorithm is confirmed on $1000$ Watts-Strogatz \cite{watts1998collective}, Erd\H{o}s-R\'enyi \cite{erdHos1960evolution}, and Barab\'asi-Albert \cite{Barabasi1999a} networks of size $N=10$, respectively, where the ground truth is obtained through an exhaustive search over all possible donation structures when each node has a single recipient on a given population structure (Supplemental Table S1).

To offer intuition in support of the optimization result, we provide an efficient approximation for the net cost paid by the cooperator relative to a random individual two steps away as
\begin{equation}
	\mathcal{C}_0 - \mathcal{C}_2 \approx c \left[\left(\frac{N}{2\xi}-1\right)\sum_i\frac{k_iI_i}{2L} + \epsilon_c\right].
	\label{C0_C2}
\end{equation}
The corresponding net benefit is:
\begin{equation}
	\mathcal{B}_0 - \mathcal{B}_2 \approx b \left(\frac{1}{2L}\sum_{i,j}I_{ij}\left[\frac{N w_{ij}}{4\xi}+k_j \left(\frac{N w_{ij}}{4\xi k_i}-1\right)\right] + \epsilon_b\right),
	\label{B0_B2}
\end{equation}
where $w_{ij} = w_{ji} =1$ if there is an edge between nodes $i$ and $j$ on the unweighted network ($w_{ij} = w_{ji} =0$ otherwise).
We have $k_i=\sum_j w_{ij}$, representing the number of neighbors (degree) of individual $i$ and $L=\frac{1}{2}\sum_k k_i$ indicates the number of undirected links in the population structure.
The heterogeneity of the network is captured by the ratio $\xi=\langle k^2 \rangle/\langle k \rangle^2$, and $\epsilon_c$ and $\epsilon_b$ are negligible compared to other terms.

When each donor has a single recipient, we have $I_i=1$ for each individual $i$, thus the net cost $\mathcal{C}_0 - \mathcal{C}_2$ does not change with different preferential recipients.
In homogenous networks ($\xi \approx 1$), the net benefit $\mathcal{B}_0-\mathcal{B}_2$ does not change with different preferential recipients since $k_i \approx k_j$ for any $i$ and $j$.
In contrast, we find that the net benefit can be increased in heterogeneous networks ($\xi \gg 1$) if an individual with degree less than $N/4\xi$ donates to a neighbor with larger degree.
To reduce the critical threshold for promoting the emergence of cooperation, one needs to increase the net benefits ($\mathcal{B}_0-\mathcal{B}_2$) of cooperators relative to the random individual two steps away, and reduce the net costs ($\mathcal{C}_0 - \mathcal{C}_2$) in the meantime.
Therefore, if a node's degree is less than $N/4\xi$, it should choose its highest-degree neighbor as its preferential recipient.
In contrast, nodes with degrees greater than $N/4\xi$ should choose the neighbor with the minimum degree as their preferential recipients.
We show our rule attains a critical ratio only slightly higher than the optimal $\mathcal{C}^*$ (Fig.~\ref{Fig3-opt}(c)), and the degree of the optimal recipients on different networks are also consistent with our rule, with $N/4\xi$ being the threshold of degree for choosing the preferential recipient (Fig.~\ref{Fig3-opt}(d)--(f)).
Note that for Erd\H{o}s-R\'enyi and Watts-Strogatz networks---wherein heterogeneity is relatively low---there are no other nodes with degrees exceeding the threshold.
In these cases, all nodes should be preferential donate to their highest degree neighbor (Fig.~\ref{Fig3-opt}(e),(f)).

\begin{figure*}[!ht]
	\centering
	\includegraphics[width=\textwidth]{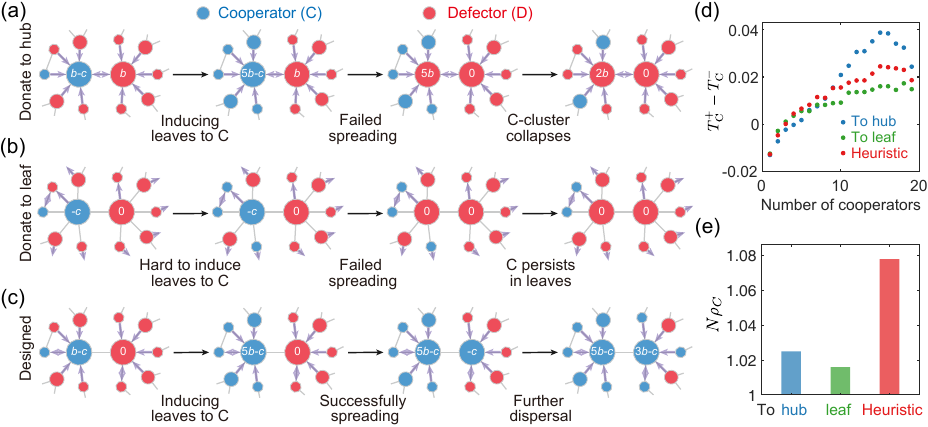}
	\caption{Mechanisms for promoting cooperation with preferred recipients.
	The stages for the fixation of cooperation includes forming local cooperative cluster, spreading cooperation to other hubs, and the further dispersal of cooperators around the hubs.
	(a) When all nodes donates to hubs, the hub becomes highly influential and drives neighbors to cooperators once it turns to cooperation.
	However, the stability of the local cooperative cluster is threatened since the cooperative hub donates to the other defective hub.
	(b) In contrast, when all nodes donate to small nodes, its hard for hubs to drive more neighbors to cooperation due to its relative low payoff.
	(c) Our heuristic rule enables small nodes donate to hubs and vice versa, which reduces the payoffs of the defective hubs while maintaining the high impact of the cooperative hubs.
	(d) We implement these three configurations on a network with $20$ nodes constructed by Barab\'asi-Albert model, and calculate the probability that the number of cooperators increases ($T_{\text{C}}^+$) or decreases by one ($T_{\text{C}}^-$).
	(e) The heuristic rule shows a high probability of the net increase in the number of cooperators over the course of evolution, which leads to the highest fixation probability of cooperation.
	}\label{Fig4-mechanism}
\end{figure*}

\subsection{Intuition behind the benefit of preferential interactions}

Both the optimization results and our proposed rule show the benefit of degree dissassortivity in the donation structure.
But how can we intuitively explain the underlying mechanism from the perspective of strategy dispersal?
Starting from a single random cooperator in a population of full defection, the survival and dispersal of cooperators is affected by the formation of a local cooperative cluster early on.
Here, a hub plays an important role in driving its low-degree neighbors to cooperation.
Figure~\ref{Fig4-mechanism}a presents the evolutionary process when all nodes donate to their largest neighbors, where the hub obtains more payoffs, hence becoming more attractive to imitate. 
This in turn drives more low-degree neighbors to cooperators.
The next step for the further dispersal of cooperators is to spread cooperation to other defecting hubs, who also receive the benefit provided by the cooperative hub.
This in turn threatens the stability of the local cooperative cluster, which will immediately collapse when the cooperative hub imitates the defective hub with a relative higher payoff among neighbors.

In contrast, when donations flows to leaves, namely the lowest-degree neighbors, the risk of disintegrating cooperative clusters when a hub defects is mitigated.
At the same time, this reduces the ability of the cooperative hubs to quickly drive cooperative clusters (Fig.~\ref{Fig4-mechanism}(b)).
Instead, our designed donation structure allows small nodes donate to hubs and vice versa. 
This enables the hubs to quickly drive their neighbors to cooperation, and more importantly, to avoid enhancing the defective hubs in the early stage (Fig.~\ref{Fig4-mechanism}(c)). 
We further confirm this mechanism by calculating the probability that the number of cooperators increases ($T_{\text{C}}^+$) or decreases ($T_{\text{C}}^-$) by one in a heterogeneous structured population during the evolution (Fig.~\ref{Fig4-mechanism}(d)).
We show that the designed configuration has a higher probability increment ($T_{\text{C}}^+ - T_{\text{C}}^-$) than donating to hubs for all nodes when there are only few cooperators in populations.
On the other hand, our heuristic also fully capitalizes on the ability of influential hubs to drive neighbors to cooperation compared to the case of donating to leaf for all nodes when the total number of cooperators increases.
Combining these two advantages, our heuristic rule results in the highest fixation probability of cooperation ($\rho_\text{C}$, Fig.~\ref{Fig4-mechanism}(e)).  

\begin{figure*}[!ht]
	\centering
	\includegraphics[width=\textwidth]{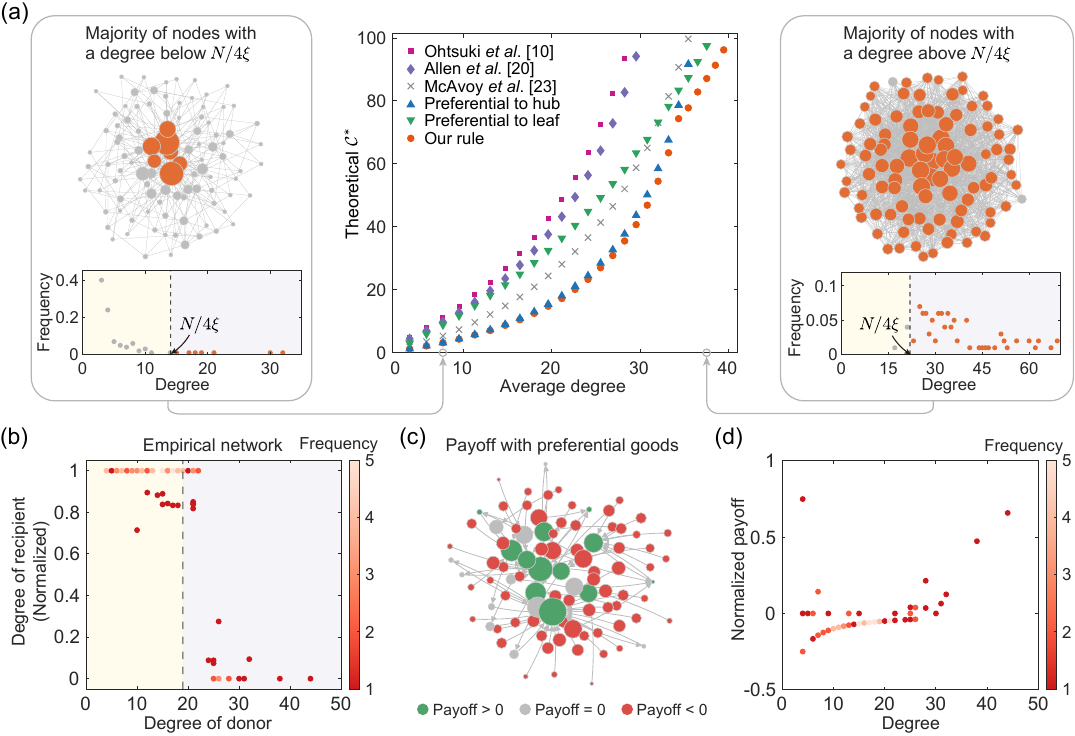}
	\caption{
		\textbf{Designing favorable preferential donation structures on empirical an scale-free synthetic networks.}
		(a) We compare the theoretical $\mathcal{C}^*$ for our heuristic rule of choosing recipients (orange dot) with other settings on scale-free networks with 100 nodes and different average degrees.
		Ohtsuki \textit{et al.} \cite{ohtsuki2006simple} (magenta square) and Allen \textit{et al.} \cite{allen2017evolutionary} (purple diamond) using the homogenous distribution of goods---cooperators benefit all neighbors, have the highest $\mathcal{C}^*$.
		The values for the fixed-cost, fixed benefit model proposed by McAvoy \textit{et al.} \cite{McAvoy2020} are marked with a gray cross.
		We also present two extreme cases of preferential social interactions---each individual choose hub (upper triangle in blue) or leaf (lower triangle in green) as its recipient.
		The network with average degrees of $6$ and $37.5$ are marked and presented in the left and right panels, respectively.
		We show that only $7\%$ nodes have degrees larger than the threshold $N/ 4\xi$ (light purple region) in our heuristic for scale-free networks with mean degree $6$, which are marked in orange, where $\xi$ captures the heterogeneity of the population structure.
		In contrast, the proportion of nodes with a degree larger than $N/ 4\xi$ in dense networks can reach $95\%$.
		(b) We optimize the single recipient for each individual with our protocol on an empirical network describing face-to-face contacts in an office building.
		The optimal preferential interaction for each individual is consistent with our rule, wherein a donor with degree below $N/ 4\xi$ (light yellow region) tends to have recipient with maximal degree, while those in the light purple region tends to give to low-degree-recipients.
		(c) We illustrate the individual payoff with optimal preferential donation structures in a full cooperation state when $b=c$, where an individual with payoff greater (smaller) than $0$ is marked in green (red), and those with payoff of $0$ is marked in gray.
		(d) We show the relation between a node's degree with its normalized payoff by degree on the undirected population structure.
		The color represents the frequency of individuals with the corresponding degree and payoff.
	} \label{Fig5-design}
\end{figure*}

\subsection{Designing the optimal donation structure on any network}
As a brief summary of the optimization results, our heuristic rule uses only a simple metric $N/4\xi$ to determine whether an individual should choose the largest or smallest neighbor.
Figure~\ref{Fig5-design}a compares of the critical ratio between our rule vs. alternatives on scale-free networks constructed by Barab\'asi-Albert model \cite{Barabasi1999a}.
We show that the traditional setting \cite{ohtsuki2006simple}---wherein each individual benefits all its neighbors---has the largest critical ratio, which hinders the cooperation most.
The model with average payoff \cite{allen2017evolutionary} has similar values to that with accumulated payoff.
The fixed cost, fixed benefit model \cite{McAvoy2020} performs between the case where individuals are preferential to hubs (the nodes with the maximum degree among the neighbors) and that with individuals preferential to leaves (the nodes with the minimum degree among the neighbors).
Here, our heuristic rule has the lowest critical ratio over networks with a range of different degrees.

\begin{figure*}[!ht]
	\centering
	\includegraphics[width=\textwidth]{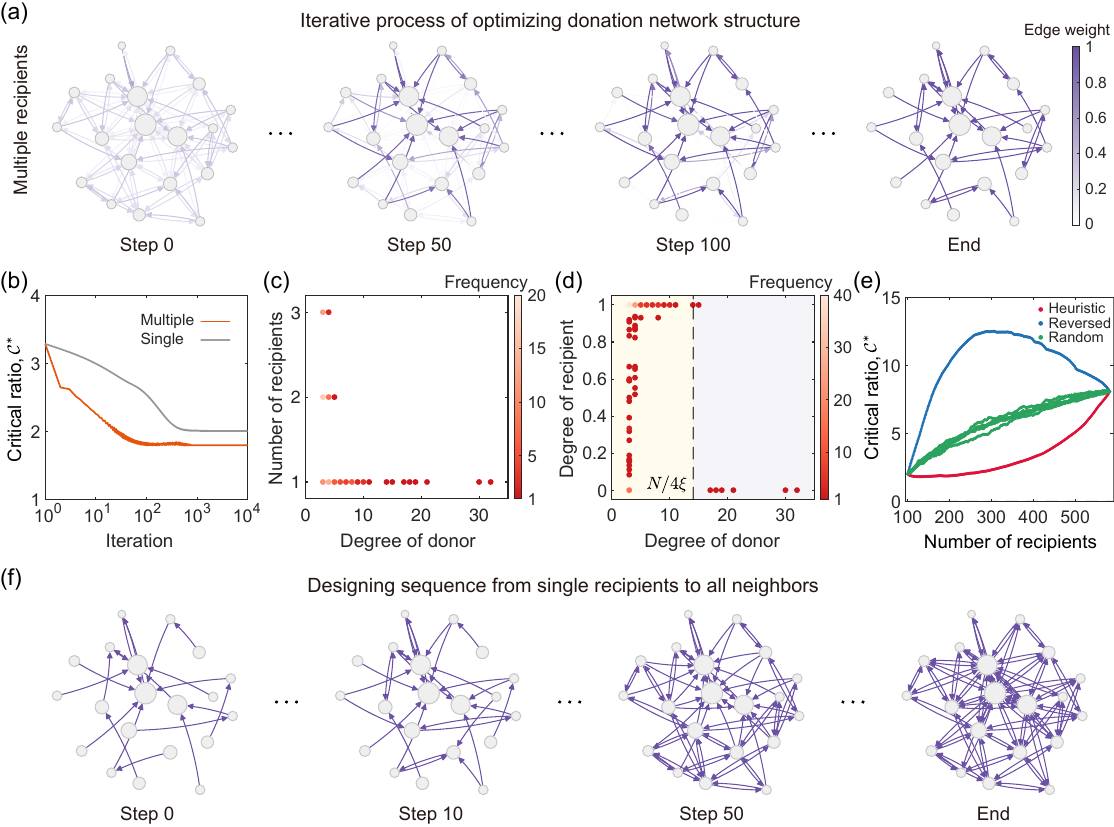}
	\caption{Optimizing multiple recipients for each individual on networks.
		(a) We illustrate the network structure of preferential social goods at iteration step $0$, $50$, $100$ and the end of the optimization process, where each node is allowed to have at least one recipient. 
		The edge weights in donation structure are marked in colors.
		(b) We present the convergence of the objective function $\mathcal{C}^*$ for scale-free networks over $10^4$ iterations of our optimization protocol.
		The number and degrees of the optimal recipients as a function of the degree of donor are shown in (c) and (d) respectively.
		The color represents the frequency of the producer-recipient pairs.
		(e) We further offer the rule for designing the trajectory from a single recipient to multiple recipients.
		(f) A sequence of producer-recipient pairs are added to the existing configuration iteratively until all individual donates to all their neighbors.
		Our heuristic rule (red dots) presents the slowest increase of $\mathcal{C}^*$.
		In contrast, the converse of our rule (blue dots) shows the steepest increase of $\mathcal{C}^*$, which may even induce the donation structure that are worse for cooperation than benefiting all neighbors.
		For comparison, we also show 10 realizations in which donor-recipient pairs are chosen uniformly at random (green dots), presenting the steady increase compared to other two cases.
	} \label{Fig6-multiple}
\end{figure*}

Interestingly, there is an intersection between the values of $\mathcal{C}^*$ of the case where individuals are preferential to hubs and that with individuals preferential to leaves (Fig.~\ref{Fig5-design}(a)).
When networks are sparse, namely the average degree is relatively low, we find that only a few nodes have degrees greater than $N/4\xi$, while a large amount of nodes with degrees less than $N/4\xi$.
Under our rule, most nodes should choose neighboring hubs as their preferential recipients, thus the case where all nodes benefit hubs have a lower critical ratio than that with all nodes preferential to leaves.
However, when networks become dense, most nodes tend to have degrees greater than $N/4\xi$, while comparatively a few nodes have degrees less than $N/4\xi$.
This explains the better performance of the case where individuals preferentially donate to leaves over hubs in dense networks.
Therefore, the intersection of $\mathcal{C}^*$---between the case that individuals donate preferentially to hubs and that to leaves---naturally appears as the average degree of networks increases, and the distribution of goods in fixed cost, fixed benefit model is in between these two extreme cases, resulting in the $\mathcal{C}^*$ lying in the middle of the corresponding values of these two cases.
Furthermore, we apply our optimization protocol on an empirical network describing contacts in an office building \cite{office2013}, and our rule also fits the optimal preferential recipient (Fig.~\ref{Fig5-design}(b)).
Figure~\ref{Fig5-design}c presents the payoff received by each individual in a full cooperation state with the optimal preferential recipients.
Furthermore, we show that the average payoffs (normalized by the node's degree) diverge at low-degree nodes, while benefits at high-degree nodes often exceed costs (Fig.~\ref{Fig5-design}(d)).

\subsection{Multiple preferential recipients}

What if we allow cooperators to donate to multiple recipients?
To answer the question, we optimize the critical ratio $\mathcal{C}^*$ using the proposed protocol with a penalty function to ensure the condition that each individual have one or more recipients, which belong to its neighbors in the population structure (see Supplementary Information), wherein each individual is allowed to have at least one preferential recipient (Fig.~\ref{Fig6-multiple}(a)).
We find that the optimal $\mathcal{C}^*$ with multiple preferential recipients may be even lower than that with single preferential recipient (Fig.~\ref{Fig6-multiple}(b)).
Nodes with low degree tend to have more recipients despite the paucity of neighbors (Fig.~\ref{Fig6-multiple}(c)), while high-degree nodes tend to have a single preferential recipient---usually the lowest degree among its neighbors (Fig.~\ref{Fig6-multiple}(d)).
We show that the degree threshold $N/4\xi$ is still consistent with the optimization results in the scenario with multiple preferential recipients.

Furthermore, we plan the trajectory toward the traditional donation structure where each individual provides social goods equally to all neighbors starting from providing social goods to a single preferential recipient.
Each step in the trajectory involves a single new producer-recipient pair relative to the existing donation structure.
According to Eqs.~(\ref{C0_C2}) and (\ref{B0_B2}), our designed trajectory of arranging the ordering of producer-recipient pairs is as follows: select node $i$ in ascending order of degree, and select the recipient $j$ in descending order of $N w_{ij}k_j /4\xi k_i -  k_j$ to maximize the net benefit at minimum cost at each step.
We find that the designed trajectory can slow down the increase in $\mathcal{C}^*$ compared to the reversed ordering of this rule and the random ordering of adding producer-recipient pairs (Fig.~\ref{Fig6-multiple}(e),(f)).

\section{Discussion}
There is a large body of research exploring the mechanisms underlying the emergence of cooperation in structured populations \cite{Su2019,McAvoy2020,li2020evolution,Zhou2021,Meng2024}, yet most have assumed that cooperators benefit their neighbors equally and allocate the social goods without preference \cite{ohtsuki2006simple,allen2017evolutionary,McAvoy2020}.
We find that the ubiquitous preferential social interactions in realistic scenarios can greatly facilitate the emergence of cooperation.
To maximize the promotion of cooperation, we develop an efficient algorithm to find the optimal distribution of social goods on any network, which can be distilled into a simple rule based on a degree threshold that captures the disassortativity in the ideal donation structure. 

A promising application for our rule is to guide the allocation of social goods in large empirical systems efficiently (Fig.~\ref{Fig5-design}(b),(c)). 
Our rule requires nodes with degree higher than the threshold $N/4\xi$ to preferentially benefit the smallest neighbor and vice versa. 
As a result, we find that the wealth is concentrated in large nodes in sparse networks, but dispersed to relatively small nodes in dense networks under optimal preferential interactions (Supplemental Fig.~S1). 
The reason is that, for dense networks, most nodes have degrees higher than the threshold $N/4\xi$, which makes almost all nodes tend to choose the smallest neighbors as their recipients (Fig.~\ref{Fig5-design}(a)).

One of the most important insights of our study into the dynamics of cooperative evolution is the incorporation of machine learning based optimization.
Many studies have uncovered the mechanisms or, in particular, the specific cases that promote the emergence of cooperation \cite{Su2022,McAvoy2020,santos2005scale,perc2010coevolutionary,Meng2024}.
However, it is natural to ask whether there is an optimal combination of different mechanisms to best facilitate cooperation.
In general, due to the nonlinear nature of evolutionary dynamics, it may be difficult to reach the optimal solution directly.
With the rise of machine learning methods, one can easily adapt the idea of designing objective functions in our study to other problems in evolutionary game dynamics.
Therefore, we expect to see the effective application through optimization methods in this area, and how it will change the paradigm of studying collective dynamics.
Our research opens the door to this exciting application by starting with optimizing the allocation of social goods in evolutionary games.

\section*{Appendix A: The critical threshold for favoring cooperation}
The population structure containing $N$ individuals is captured by a undirected network, where nodes indicate individuals and edges represent who may imitate the strategy from whom.
The interactions between individuals are captured by a donation structure, where a directed edge from $i$ to $j$ indicates that $I_{ij}=1$ and $i$ will donate a benefit to $j$ when it is a cooperator. Otherwise $I_{ij}=0$ and there is no directed edge from $i$ to $j$.
Cooperators interact with their neighbors and provide the benefit to the preferential recipients according to the donation structure in each round of game.

The state of the population can be represented by a binary vector $\mathbf{x}\in \{0,1\}^N$, with $x_i=1$ denoting individual choosing cooperation and $x_i=0$ indicating defection.
We then obtain the accumulated payoff for individual $i$ in state $\mathbf{x}$ given by $f_i(\mathbf{x}) = -cI_ix_i+b\sum_{j=1}^{N}I_{ji}x_j$, where $I_i=\sum_{j=1}^{N}I_{ij}$ represent the number of recipient for individual $i$.
After each game round, an individual $i$ is uniformly at random chosen to update its strategy by imitating the strategy from its neighbor $j$ with probability proportional to $F_j(\mathbf{x})$.
Therefore, the probability of $j$ transmitting its strategy to $i$ in state $\mathbf{x}$ is $r_{ji}(\mathbf{x})=\frac{1}{N} \frac{w_{ij} F_j(\mathbf{x})}{\sum_{k=1}^{N} w_{ik}F_l(\mathbf{x})}$.
And the fixation probability of cooperation is 
\begin{equation}
\begin{aligned}
	\rho_{C}=\frac{1}{N}+\delta &\sum_{t=0}^{\infty}\sum_{\mathbf{x}\in \{0,1\}^N} \mathbb{P}_{\mathbf{u}}^{\circ}\left[\mathbf{X}(t)=\mathbf{x} \right]  \sum_{i =1}^{N} \\ &\pi_i\sum_{j=1}^{N}(x_j-x_i)\left.\frac{\mathrm{d} r_{ji}(\mathbf{x})}{\mathrm{d} \delta}\right|_{\delta=0} +\mathcal{O}\left(\delta^{2}\right),\nonumber
\end{aligned}
\end{equation}
where $\mathbb{P}_{\mathbf{u}}^{\circ}\left[\mathbf{X}(t)=\mathbf{x} \right]$ indicates the neutral probability of the system reaching state $\mathbf{x}$ at time step $t$ starting from a single random cooperator (see details in Supplementary Note 1).
By defining $\eta_{ij} :=\sum_{t=0}^{\infty}\sum_{\mathbf{x}\in \{0,1\}^N} \mathbb{P}_{\mathbf{u}}^{\circ}\left[\mathbf{X}(t)=\mathbf{x} \right] \left(\hat{x}-x_i x_j\right)$ where $\widehat{x}$ denotes the reproductive-value-weighted frequency \cite{mcavoy2021fixation} of cooperation in state $\mathbf{x}$, we obtain the exact formula of fixation probability
\begin{equation}
	\begin{aligned}
		\rho_{C}= &\frac{1}{N} 
		- \frac{\delta c}{N} \sum_{i,j=1}^N \pi_i p_{ij}^{(2)} I_j \eta_{ij} + \\ & \frac{\delta b}{N} \left( \sum_{i,j,k=1}  \pi_i p_{ij}^{(2)}I_{kj} \eta_{ik}-\sum_{i,j=1}^{N} \pi_i I_{ji} \eta_{ij} \right)
		+O\left(\delta^{2}\right), \nonumber
	\end{aligned}
\end{equation}
where $\eta_{ij}$ is the unique solution of the equations
\begin{equation}
	\eta_{ij}= \begin{cases} \frac{1}{2} +\frac{1}{2}\sum_{k=1}^N p_{ik} \eta_{kj} + \frac{1}{2}\sum_{k=1}^N p_{jk} \eta_{ki} & \text{if}~~i \neq j \\ 0 & \text{if}~~i=j\end{cases}.
	\label{eta_ij}
\end{equation}
Therefore, cooperation is favored over defection when $b/c > \mathcal{C}^*$, and Eq. (\ref{exact})
is obtained.

\section*{Appendix B: Optimization of preferential allocations}
To find the optimal preferential recipient(s) for each individual, we design the penalty functions for the case of a single recipient and that of multiple recipients respectively.
For the optimization on the donation structure where each individual is allowed to have a single recipient, namely $I_i=1$ for all $i$, we define the penalty function
\begin{equation}
	\begin{aligned}
		P_s =& (\mathcal{C}^*)^2 + \sigma \sum_{i=1}^N \left( \sum_{k=1}^N I_{ik} -1 \right)^2 + \sigma \sum_{i,j=1}^N I_{ij} (1-w_{ij}) \\ &+ \gamma \sum_{i,j=1}^N I_{ij} (1-I_{ij}), \nonumber
	\end{aligned}
\end{equation}
where $\sigma$ and $\gamma$ are penalty factors, with $\sigma_{k+1} = \rho \sigma_{k}$ and $\gamma_{k+1} = \rho \gamma_{k}$  at each step of iteration.
Here we set $\rho=1.01$, $\sigma_0=1$ and $\sigma_k \leq 1\times 10^4$,  $\gamma_0=0.01$ and $\gamma_k \leq 100$.
We define $I_{ij} := 1/\left(1+ \text{exp}(-\theta_{ij})\right)$, which naturally leads to $I_{ij}\in (0,1)$. This ensures that the latter two terms in the penalty function also have positive values. And we can obtain the gradient of $\frac{\partial P_s}{ \partial \theta_{ij}}$ by
\begin{equation}
	\begin{aligned}
		\frac{\partial P_s}{ \partial \theta_{ij}} =& 2 \mathcal{C}^* \frac{\partial \mathcal{C^*}}{\partial I_{ij}} \frac{\partial I_{ij}}{\partial \theta_{ij}} + 2 \sigma  \left( \sum_{k=1}^N I_{ik} -1 \right) + \sigma(1-w_{ij})\frac{\partial I_{ij}}{\partial \theta_{ij}} \\ & + \gamma (1-2 I_{ij}) \frac{\partial I_{ij}}{\partial \theta_{ij}}, \nonumber
	\end{aligned}
\end{equation}
where $\frac{\partial \mathcal{C^*}}{\partial I_{ij}}$ can be calculated by solving a system of $N(N-1)/2$ linear equations based on the recurrence relationship in Eq.~(\ref{eta_ij}).
We apply the RMSprop \cite{tieleman2012lecture} algorithm with gradient descent at each step of iteration to minimize $P_s$.
The process starts from $I_{ij}=I_{ji}=w_{ij}/w_i$ for each node---where each individual divides a benefit equally among its neighbors---with learning rate of $0.1$, decay rate of $0.9$ and ends when $ | \Delta P_s|<10^{-5}$.

For multiple preferential recipients, each individual is allowed to have $I_i \geq 1$ number of recipients.
Therefore, we define the penalty function 
\begin{equation}
	\begin{aligned}
		P_m(\boldsymbol{\theta}) &= (\mathcal{C}^*)^2 + \sigma \sum_{i=1}^N \mathbbm{1}_{\{I_i<1\}} \left( \sum_{k=1}^N I_{ik} -1 \right)^2 \\ &+ \sigma \sum_{i,j=1}^N I_{ij} (1-w_{ij})+\gamma \sum_{i,j=1}^N I_{ij} (1-I_{ij}), \nonumber
	\end{aligned}
\end{equation}
where the indicator function $\mathbbm{1}_{\{I_i<1\}}$ is equal to $1$ if $I_i<1$ otherwise equal to $0$.
The gradient of $P_m$ to $\theta_{ij}$ is then given by
\begin{equation}
	\frac{\partial P_m}{ \partial \theta_{ij}} = \begin{cases} 2 \mathcal{C}^* \frac{\partial \mathcal{C^*}}{\partial I_{ij}} \frac{\partial I_{ij}}{\partial \theta_{ij}} + 2 \sigma  \left( \sum_{k=1}^N I_{ik} -1 \right)\frac{\partial I_{ij}}{\partial \theta_{ij}} \\ + \sigma (1-w_{ij})\frac{\partial I_{ij}}{\partial \theta_{ij}} + \gamma (1-2 I_{ij}) \frac{\partial I_{ij}}{\partial \theta_{ij}}& \text{if}~I_i<1, \\ 2 \mathcal{C}^* \frac{\partial \mathcal{C^*}}{\partial I_{ij}} \frac{\partial I_{ij}}{\partial \theta_{ij}}  + \sigma (1-w_{ij})\frac{\partial I_{ij}}{\partial \theta_{ij}} \\ + \gamma (1-2 I_{ij}) \frac{\partial I_{ij}}{\partial \theta_{ij}} & \text{if}~I_i \geq 1.\end{cases} \nonumber
\end{equation}
The other parameters during the implementation of the optimization procedure are the same as the case of a single recipient.
 
%

\end{document}